\documentclass[12pt]{iopart}
 
\expandafter\let\csname equation*\endcsname\relax
\expandafter\let\csname endequation*\endcsname\relax
\usepackage{amsmath,amssymb,mathtools}

\usepackage{graphicx,color}
\usepackage{enumerate}

\newcommand\ra{\rangle}
\newcommand\be{\begin{equation}} 
\newcommand\ee{\end{equation}} 
\begin{document}

\title{Unusual equilibration of a particle in a potential with a thermal wall}

\author{Deepak Bhat$^1$, Sanjib Sabhapandit$^2$  Anupam Kundu$^1$, and Abhishek Dhar$^1$}  
\address{$^1$International Centre for Theoretical Sciences, TIFR, Bangalore - 560089, India\\
$^2$Raman Research Institute,  Bangalore - 560080, India}

\date{\today}

\begin{abstract}
We consider a particle in a one-dimensional box of length $L$ with a
Maxwell bath at one end and a reflecting wall at the other end. Using
a renewal approach, as well as directly solving the master equation,
we show that the system exhibits a slow power law relaxation with a
logarithmic correction towards the final equilibrium state.  We extend
the renewal approach to a class of confining potentials of the form
$U(x) \propto x^\alpha$, $x>0$, where we find that the relaxation is
$\sim t^{-(\alpha+2)/(\alpha-2)}$ for $\alpha >2$, with a logarithmic
correction when $(\alpha+2)/(\alpha-2)$ is an integer. For $\alpha <2$
the relaxation is exponential. Interestingly for $\alpha=2$ (harmonic
potential) the localised bath can not equilibrate the particle.

\end{abstract}
\maketitle

\section{Introduction}

The question of approach to the equilibrium of a system connected to a thermal bath is a well-known and important problem. For a colloidal particle in a thermal environment, it is known that the velocity (and position, for a particle in a generic confining potential) equilibrates to the Maxwell-Boltzmann distribution and the relaxation occurs exponentially in time. However, there are many
examples where a slower non-exponential relaxation is observed, such as in sub-diffusive processes~\cite{barkai1999}, glassy dynamics,
systems with initial condition drawn from heavy-tailed distributions~\cite{toenjes2013}, processes with resetting~\cite{sabhapandit2015}, reversible diffusion-controlled chemical reactions \cite{Szabo,Huppert,Gopich,Pines}, and diffusion in a logarithmic potential~\cite{mukamel2011}. In the present work, we consider another
example where the system is connected to a heat bath localised in space. For this system we observe slow relaxation. Note that when the particle is connected to extended baths like Langevin baths, the interaction with the bath is present at all times at every positions and that usually results in exponential relaxation \cite{Risken,Van_kampen}.   Although localised baths have been used in energy transport problems \cite{AD} and in driven granular systems such as a ball bouncing on a vibrating plate under gravity \cite{Chastaing, Luck}, the approach to the stationary state has not been studied.

Our model consists of a particle of mass $m$, moving inside a one-dimensional box of size $L$. Inside the box, the particle moves
ballistically and gets elastically reflected at the right wall at the position $x=L$. On the left wall at position $x=0$, the particle
interacts with a thermal reservoir taken to be Maxwell bath. On every collision with the bath, the particle emerges with a new velocity
$u>0$, chosen from the distribution $f(u)=\beta m u e^{-\beta m u ^2/2}$, where $\beta$ is the inverse temperature \cite{Salwen}. Without any loss of
generality, we set $\beta=1$ and $m=1$ throughout the paper. It is well-known and easy to verify that the steady state of this system is the Maxwell-Boltzmann distribution. However, the approach to this equilibrium state has not been studied earlier.

\begin{figure}
\centering
 \includegraphics[height=12cm,angle=-90,keepaspectratio=true]{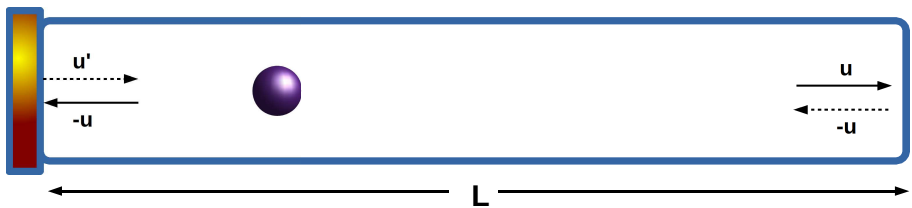} \caption{\label{particle-in-box}
 A point particle moves between a thermal reservoir kept at $x=0$ and a reflecting wall at $x=L$. Whenever particle approaches the
 reservoir it renews its velocity and comes with velocity chosen from Rayleigh distribution while it gets reflected once it approaches the wall. Arrows with solid and dashed lines respectively indicate the velocity of the particle  before and after collision, either with the reservoir on the left end or with the reflecting wall on right end of the cylinder.}
\end{figure}

In this paper we explicitly compute the propagator corresponding to the Markov process and show that the relaxation to the equilibrium
state is non-exponential. In particular, we find an exact solution of the master equation in the Laplace domain, from which we find that the joint distribution $P(x,u,t)$ of the position $x$ and the velocity $u$ of the particle approaches to equilibrium
Maxwell-Boltzmann distribution as 
\begin{equation}
\left [P(x,u,t) - P_{MB}(x,u) \right ]_{t \to \infty} \sim \frac{\log (t)}{t}~, \label{relax-log}
 \end{equation} 
where $P_{MB}(x,u)=\frac{\theta(x)\theta(L-x)}{L}\frac{e^{-u^2/2}}{\sqrt{2 \pi  }}$ is the equilibrium distribution.  The exact solution is obtained by directly solving the master equation, and also, by using the renewal property of the dynamics.

We also consider the situation where, instead of a box of finite size, the particle is confined in a potential of the form $U(x)\propto x^\alpha$ for $0\leq x < \infty$ and $0<\alpha<\infty$. In this case, we can generalise the renewal approach and find that the relaxation towards equilibrium depends explicitly on the value of $\alpha$ as :
\begin{equation}
\left [P(x,u,t) - P_{MB}(x,u) \right ]_{t \to \infty} \sim 
\begin{cases}
& {t^{-(\alpha+2)/(\alpha-2)}},~~~~\text{for}~\alpha >2 \\
& \text{no~relaxation},~~\text{for}~\alpha =2 \\
& \text{exponential},~~\text{for}~\alpha <2
\end{cases},
\end{equation}
with a logarithmic correction for positive integer values of
$\frac{\alpha+2}{\alpha -2}$ in the $\alpha>2$ case.

The paper is organised as follows: In Sec. \eqref{ME} we construct a
master equation for the particle moving in between the reservoir and a
reflecting wall and, obtain the solution to the master equation in
Laplace space.  In Sec.\eqref{Renewal}, we re-drive this solution using renewal
property of the process. In order to study the question of relaxation
to equilibrium, in Sec.\eqref{Approach} we analyse the solution in the appropriate
limit of the Laplace variable and find that the relaxation is power
law with logarithmic correction.  In Sec.\eqref{General} we generalise this problem
of relaxation for particle confined in a potential instead of a box
and find that the approach to the equilibrium now depends on the form
of the potential. Finally in Sec.\eqref{Conclusion}, we provide concluding remarks.

\section{Master equation}
\label{ME}
\noindent
The master equation describing the time evolution of the phase-space
distribution $P(x,u,t)$ is given by
\begin{eqnarray}\fl\qquad
\frac{\partial P(x,u,t)}{\partial t} &= \mathcal{L}P(x,u,t)  \label{mastereq1}\\
\fl\text{with,}\nonumber\\
\fl\qquad\mathcal{L}P(x,u,t) & =-u\frac{\partial P}{\partial x} + \delta(x) \left[- f(u) \int u' \theta(-u')P(x,u')du' + u\theta(-u)~P(x,u)\right]  \nonumber \\
\fl\qquad &\quad~ - \delta(L-x) u\left[\theta(-u) ~ P\left(x, -u \right) +
~ \theta\left(u\right) ~P(x,u) \right] \label{mastereq2}~
\end{eqnarray}
where the first term corresponds to the ballistic motion, second part to collisions with the thermal wall and the last term corresponds to collisions with the right wall. The general solution of this equation can be written as
\begin{equation}
P(x,u,t)= \int_0^L dx_0 \int_{-\infty}^{\infty} du_0 G(x,u,t|x_0,u_0,0) P_0(x_0,u_0)~, \label{equationpg}
\end{equation}
where $G(x,u,t|x_0,u_0,0)$ is the propagator and given by the solution of \eref{mastereq1} with the initial condition $P(x,u,t=0)=\delta(x-x_0)\delta(u-u_0)$ and $P_0(x_0,u_0)$ is the distribution of the initial condition. Therefore, $G(x,u,t|x_0,u_0,0)$ must satisfy :
\begin{eqnarray}
 \frac{\partial G}{\partial t} - \mathcal{L}G=\delta(t)\delta(x-x_0)\delta(u-u_0) \label{mastereq3}
\end{eqnarray}
Taking a Laplace transform with respect to time gives 
\begin{eqnarray}
 s \tilde{G}(x,u,s|x_0,u_0) -\delta(x-x_0)\delta(u-u_0) = \mathcal{L}\tilde{G}(x,u,s|x_0,u_0)\label{me-lap}
\end{eqnarray}
where $ \tilde{G}(x,u,s|x_0,u_0)=\int^{\infty}_0 e^{-st} G(x,u,t|x_0,u_0,0)dt$. Away from the boundaries the collision terms in \eqref{mastereq2} drop and we get
\begin{eqnarray}
\frac{\partial \tilde{G}}{\partial x}+  \frac{s}{u} \tilde{G}_0   = \frac{\delta(x-x_0)\delta(u-u_0)}{u}\nonumber
\end{eqnarray}
This has a general solution of the form
\begin{eqnarray}
 \fl\qquad\tilde{G}(x,u,s|x_0,u_0)=\left[ A~e^{-\frac{sx}{u}}
 +  \frac{\theta(x-x_0)\delta(u-u_0)}{u_0}\, e^{-\frac{s(x-x_0)}{u_0}}\right]\theta(x)\theta(L-x). \label{G-tilde-1}
\end{eqnarray}
where $A$ is some function of $L$, $u$ and $s$. We find that $A$ has the  form, 
\begin{eqnarray}
 A=\theta(u)A_+(u,s) +\theta(-u)A_-(u,s),
\end{eqnarray}
where $A_\pm$ are obtained from the boundary equations at $x=0$ and $x=L$, as
\begin{eqnarray}
A_+(u,s)= \frac{\bigg[\theta(u_0) e^{-\frac{s(2L-x_0)}{u_0}} + \theta(-u_0)e^{\frac{s}{u_0}x_0} \bigg]}{\left[1-\int^{\infty}_0 u' e^{-\frac{u'^2}{2}-\frac{2sL}{u'}}du' \right]} e^{-\frac{u^2}{2}}
\end{eqnarray}
and
\begin{eqnarray}
  A_-(u,s) =  A_+(-u,s)e^{\frac{2sL}{u}} + \left[ \frac{\delta(u+u_0)}{u_0}e^{-\frac{s(2L-x_0)}{u_0}} - \frac{\delta(u-u_0)}{u_0} e^{\frac{sx_0}{u_0}}\right].
\end{eqnarray}
Inserting these forms of $A_{\pm}$ in \eqref{G-tilde-1} and performing some simplifications we obtain the following explicit expression of the Green function in the Laplace space 
\begin{eqnarray}
\fl \tilde{G}(x,u,s|x_0,u_0)=\theta(x)\theta(L-x)\left[\frac{\theta(x-x_0)\theta(u_0)}{u_0}-  \frac{\theta(x_0-x)\theta(-u_0)}{u_0}\right]  \delta(u-u_0)e^{-\frac{s(x-x_0)}{u_0}} \nonumber\\
 \fl\qquad+\theta(x)\theta(L-x)\frac{\theta(u_0)\delta(u+u_0)}{u_0}e^{-\frac{s(2L-x_0-x)}{u_0}}\nonumber\\
\fl\qquad +\theta(x)\theta(L-x) \frac{\bigg[\theta(u_0) e^{-\frac{s(2L-x_0)}{u_0}} + \theta(-u_0)e^{\frac{sx_0}{u_0}} \bigg]\left[\theta(u)e^{-\frac{sx}{u}}+\theta(-u)e^{\frac{s(2L-x)}{u}}\right]}{\left[1-\int^{\infty}_{0} u' e^{-\frac{u'^2}{2}-\frac{2sL}{u'}}du' \right]}  e^{-\frac{u^2}{2}}.~~~  \label{G1}
\end{eqnarray}
We will later see that in the $s\to 0$ limit this Green's function
approaches to the form $P_{MB}(x,u)/s$. Performing inverse Laplace
transform provides $G(x,u,t|x_0,u_0,0)$, using which
in \eqref{equationpg} one can get the relaxation to the equilibrium
state. Before going into that, let us show in the next section that
the process has a nice renewal property and using which also one can
obtain \eqref{G1}.

\section{Solution using renewal equation approach} 
\label{Renewal}
\begin{figure}
\centering
 \includegraphics[height=14cm,angle=-90,keepaspectratio=true]{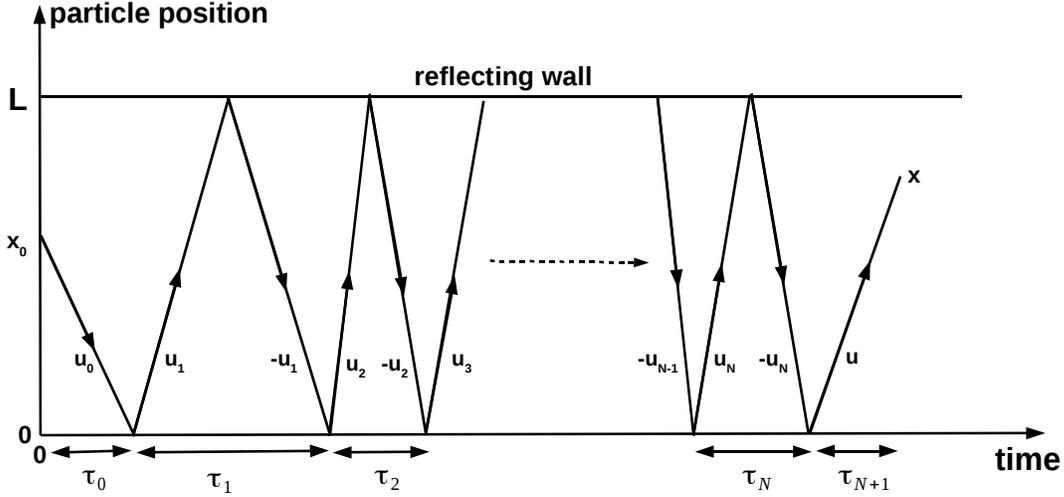} \caption{\label{renewal}
 A particle starting at $x_0$ with velocity $u_0$, moves between reflecting wall
 at $x=L$ and reservoir at $x=0$ several times before it reaches the
 final position $x$ with velocity $u$. It renews its velocity every time it
 emerges from the reservoir and simply changes sign on  reflection from the 
 wall.}
\end{figure}

In Fig. \ref{renewal}, we show a typical trajectory of the particle, in 
which one can observe that it consists of independent segments, and this 
enables us use the renewal approach. This approach is about summing over 
all the trajectories which started at $(x_0,u_0)$, terminating at $(x,u)$ 
in given time $t$. It uses a novel property that the particle renews its 
velocity whenever it is reflected from the reservoir. Let us define the 
following probabilities: 
\begin{enumerate}
\item  $H_0(x,u,\tau|x_0,u_0)$ is the probability distribution that particle 
starting from $(x_0,u_0)$ reaches $(x,u)$ in time $\tau$ without hitting 
the reservoir even for a single time. 
\item  $H(x,u,\tau)=\int_0^\infty H_0(x,u,\tau|0,u_0) f(u_0)du_0$, is the 
probability distribution for the particle to be at $(x,u)$ in time $\tau$, 
given that it started from the reservoir ($x=0$) with a velocity chosen 
from Rayleigh distribution, and has not hit the reservoir afterwards.
\item  $I_0(\tau|x_0,u_0)$ is the probability distribution that the particle 
starting from $(x_0,u_0)$ hits the reservoir at time $\tau$ for the
first time.
\item  $I(\tau)$ is the probability distribution that the particle emerging 
from the reservoir returns to it at time $\tau$. 
\end{enumerate}
In terms of these probabilities the propagator $G(x,u, t|x_0,u_0)$ can be written as
\begin{eqnarray}
\fl G(x,u,t|x_0,u_0)=H_0(x,u,t|x_0,u_0)\nonumber \\
\fl\quad~  +\sum^{\infty}_{N=0}\int^{\infty}_{0}
d\tau_0...\int^{\infty}_{0} d\tau_{N+1} I_0(\tau_0|x_0,u_0)
I(\tau_1)...I(\tau_N)H(x,u,\tau_{N+1})\, 
\delta\left(\sum^{N+1}_{i=0}\tau_i-t\right)
\label{aconvolution} 
\end{eqnarray}
where the first term $H_0(x,u,t|x_0,u_0)$ represents the probability
that the particle, starting from $(x_0,u_0)$ reaches $(x,u)$ at time
$t$ without ever hitting the reservoir. The second term contains the
contributions from other configurations where the particle reaches
$(x,u)$ at time $t$ after having hit the reservoir $N (=
1,2,...\infty)$ times. Let the collisions with the reservoir occur at times
$\tau_0,~\tau_1,~...,\tau_N$, while $\tau_{N+1}$ denotes the time
duration of last segment in which the particle comes out from the last
collision with the reservoir and reaches $(x,u)$ directly.  The delta
function ensures that the total time is $t=\sum_{i=0}^{N+1}\tau_i$.

In order to proceed, we take the Laplace transform of Eq.\eqref{aconvolution}.  
After some simplifications, we get
\begin{eqnarray}
\tilde{G}(x,u,s|x_0,u_0)
=\tilde{H}_0(x,u,s|x_0,u_0)+\frac{\tilde{H}(x,u,s)\tilde{I}_0(s|x_0,u_0)}{1-\tilde{I}(s)}, \label{mastereq-la}
\end{eqnarray}
where the tilde $\tilde{\ }$ over a function denotes the Laplace transform of
the corresponding function (in the time domain) with respect to
$\tau$.  Let us now determine $\tilde{H}_0(x,u,s|x_0,u_0)$,
$\tilde{I}_0(s|x_0,u_0)$, $\tilde{I}(s)$ and $\tilde{H}(x,u,s)$.

The probability $H_0(x,u,\tau|x_0,u_0)$ gets contributions only form 
those trajectories in which the particle reaches directly the $(x,u)$ without 
hitting the reservoir and this can happen in several ways. For instance if 
$x>x_0$, the particle, without hitting the reservoir can reach $x$  only if 
$u_0>0$. On the other hand if $x<x_0$, the particle reach $x$ directly if 
$u_0<0$ or after getting reflected back from the wall at $x=L$ if $u_0>0$. 
This gives
 \begin{eqnarray}
 \fl\quad  H_0(x,u,\tau|x_0,u_0)& =
 [\theta(x-x_0)\theta(u_0)+\theta(x_0-x)\theta(-u_0)]\nonumber \\
 &\quad\times \theta(x_0+u_0\tau)\theta(L-x_0-u_0\tau)\delta(x-x_0-u_0\tau)\delta(u-u_0) \nonumber \\
 &
 \fl +\theta(u_0)\theta(x_0+u_0\tau-L)\theta(2L-x_0-u_0\tau)
 \delta(x-2L+x_0+u_0\tau)\delta(u+u_0), \label{H0}
\end{eqnarray}
from which one gets $H(x,u,\tau)=\int_0^\infty H_0(x,u,\tau|0,u_0) f(u_0)du_0$.
Taking Laplace transform of this, we get 
\begin{eqnarray}\fl
\tilde{H}_0(x,u,s|x_0,u_0)
=\theta(x)\theta(L-x)\bigg\{\left[\frac{\theta(x-x_0)\theta(u_0)}{u_0}-\frac{\theta(x_0-x)\theta(-u_0)}{u_0}\right]
\delta(u-u_0)e^{-\frac{s(x-x_0)}{u_0}}
\nonumber \\
\qquad\qquad\qquad \qquad\qquad\qquad \qquad\qquad
+\frac{\theta(u_0)\delta(u+u_0)}{u_0}e^{-\frac{s(2L-x-x_0)}{u_0}}\bigg\}, \label{H0-2}
\end{eqnarray}
and
\begin{eqnarray}
\tilde{H}(x,u,s)=\theta(x)\theta(L-x)\bigg[\theta(u)e^{-\frac{x}{u}s}  +\theta(-u) e^{\frac{(2L-x)}{u}s} \bigg]e^{-\frac{u^2}{2}}. \label{H}
\end{eqnarray}

The probability $ I_0(\tau|x_0,u_0)$ to reach the reservoir for the
first time gets contribution from two processes: (a) if $u_0>0$ it
approaches the wall, gets reflected back and then reaches the
reservoir in time $\frac{L-x_0}{u_0}+\frac{L}{u_0}$, (b) if $u_0<0$ it
directly arrives at the reservoir in time $\frac{x_0}{-u_0}$. The
probability to reach the reservoir in time $\tau$ is then given by:
\begin{eqnarray}
 I_0(\tau|x_0,u_0)\equiv \theta(u_0)\delta\left[\tau-\left(\frac{2L-x_0}{u_0}\right) \right]  + \theta(-u_0)\delta\left(\tau+\frac{x_0}{u_0} \right). \nonumber 
 \end{eqnarray}
 It is easy to observe that $I(\tau)=\int_0^\infty I_0(\tau|x=0,u)f(u)du$. 
 Once again taking Laplace transform we get
\begin{eqnarray}
\tilde{I}_0(s|x_0,u_0) =\bigg[\theta(u_0)e^{-\frac{s(2L-x_0)}{u_0}} + \theta(-u_0)e^{\frac{sx_0}{u_0}}\bigg], 
\label{I0}
\end{eqnarray}
and
\begin{eqnarray}
\tilde{I}(s) =\int^{\infty}_{0} e^{-\frac{2sL}{u}} f(u) du =\int^{\infty}_{0} u e^{-\frac{u^2}{2}-\frac{2sL}{u}} du.
\label{I}
\end{eqnarray}
Note that the  Laplace transform $\tilde{I}(s)$ can be easily inverted 
to get the return time distribution $I(\tau)$ in each renewal
event, which has a power law tail $I(\tau)=(4L^2/\tau^3)~\exp(-2L^2/\tau^2)$.

Inserting the above expressions \eref{H0}--\eref{I}  in Eq.\eqref{mastereq-la} 
and rearranging appropriately we get Eq. \eqref{G1}, which we analyse in the 
next to understand the approach to equilibrium.

\section{Approach to equilibrium}
\label{Approach}
The particle inside the confining box must relax to the equilibrium in
the large time limit $t\to \infty$, which corresponds to the $s \to 0$
limit in the Laplace domain. In this limit, contributions from those
trajectories which are not at all approaching the reservoir are
negligible. Hence it is enough to look at the small $s$ behaviour of
the second term on the right hand side of Eq.~\eqref{mastereq-la} [or
the last term on the right hand side of Eq. \eqref{G1}]. Note that $$ \left[1-\int^{\infty}_{0} u'
e^{-\frac{u'^2}{2}-\frac{2sL}{u'}} du'\right]^{-1}= \left(\sqrt{2 \pi
} s L\right)^{-1} - \pi^{-1}\log \left(sL\right)
+ \mathcal{O}[s \log(s)].$$  Using this,
one can easily see that,
$\tilde{G}_0(x,u,s|x_0,u_0) \sim \frac{P_{MB}(x,u)}{s}$ in the leading
order, where $P_{MB}(x,u)$ is given after Eq.~\eqref{relax-log}. It is
perhaps a bit counter intuitive that the steady state distribution
becomes Maxwell-Boltzmann even if the particles emerge upon collision
with velocities chosen from Rayleigh distribution. But this can easily
be understood from particle flux balance.

To understand at the approach towards $P_{MB}(x,u)$, one needs to look
at the next order term in $s$ which is given by
\begin{eqnarray}\fl\qquad
\left[ \tilde{G}_0(x,u,s|x_0,u_0)-\frac{1}{s}P_{MB}(x,u) \right]_{s \to 0}
&\approx 
-\sqrt{\frac{2}{\pi}} \log \left(sL\right)\theta(x)\theta(L-x)\frac{e^{-\frac{u^2}{2}}}{\sqrt{2 \pi }}.~~~~~~\label{relax}
\end{eqnarray}
This immediately suggests that the approach towards equilibrium is 
\begin{equation}
\left [P(x,u,t) - P_{MB}(x,u) \right ]_{t \to \infty} \sim \frac{\log (t)}{t}~. \label{relax-log-1}
 \end{equation}
This slow power law relaxation has its root in the power law tail of the distribution $I(\tau)$ of the return times $\tau$ in each renewal events.

\begin{figure}
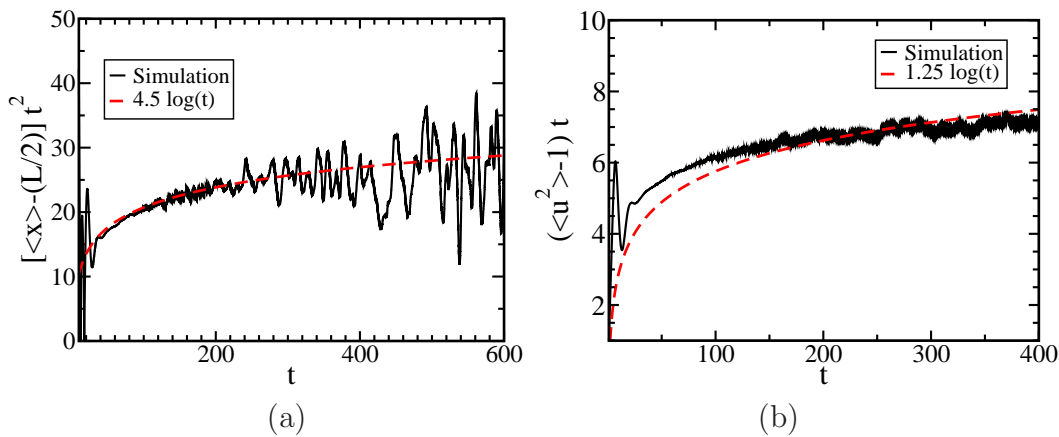

\centering
\includegraphics[height=5cm,keepaspectratio=true]{fig3a.eps}
\includegraphics[height=5cm,keepaspectratio=true]{fig3b.eps}\\
(a)\hspace{6cm}(b)
\caption{Particle in a box: behavior of average position $\langle x \rangle$ and mean squared velocity $\langle u^2 \rangle$ of the particle as a function of time obtained from simulations is shown in (a) and (b) respectively. The relaxation observed in simulations (continuous black line) agrees with the  analytic 
prediction from Eq.\eqref{G3} (dashed red line).}
\label{figalpha1}
\end{figure}

We now test our prediction from numerical simulations. Instead of
looking at the full distribution we look at how the moments relax to
their equilibrium values. In particular we look at $\langle x \rangle$
and $\langle u^2 \rangle$ as functions of time starting from initial
distribution $P(x,u,t=0)= \delta(x)f(u)$. From a similar calculation
as above one can show that
\begin{eqnarray}\fl\quad
\tilde{\langle x\rangle}&=&
\frac{\int^{\infty}_{0}u^2e^{-\frac{u^2}{2}}\left[1-e^{-\frac{sL}{u}}\right]^2du }{s^2\left[1-\int^{\infty}_{0} u' e^{-\frac{u'^2}{2}-\frac{2sL}{u'}} du'\right]}
\approx \frac{L}{2 s } - \frac{ L^3[\log(2)-1] }{3\pi } s\log  \left(sL\right) + o(s\log(s))
\end{eqnarray}
and 
\begin{eqnarray}\fl\qquad
\tilde{\langle u^2 \rangle}&=&\frac{\int^{\infty}_{0} u^3 e^{-\frac{u^2}{2}} \left[1-e^{-\frac{2Ls}{u}}\right]du}{s\left[1-\int^{\infty}_{0} u' e^{-\frac{u'^2}{2}-\frac{2sL}{u'}} du'\right]}
\approx \frac{1}{s}-L\sqrt{\frac{2}{\pi}}\log(sL) + o(\log(s))
\end{eqnarray}
where $o(a)$ denotes terms smaller than $O(a)$. This leads to   the following relaxation of the mean position and mean squared velocity to their equilibrium values ($L/2$ for the  mean position and 1 for the mean squared velocity)
\begin{eqnarray}
\left[\langle x\rangle - \frac{L}{2
} \right] \sim     \frac{ \log \left(t\right) }{t^2} ~~~\text{and} ~~~
\left[\langle u^2 \rangle - 1 \right] \sim \frac{\log(t)}{t}
\label{G3}
\end{eqnarray}
at large times. In Fig.\ref{figalpha1}, we have compared mean position and means squared velocity obtained in computer simulation with the analytical expressions and observe a  good agreement between the two.

\section{Particle in general potential}
\label{General}
In the previous section we have studied the relaxation of a single
particle confined inside a hard box of size $L$. The renewal property
of the motion of the particle was very useful in computing this
relaxation.  Interestingly, it turns out that this renewal property
can also be used to study the relaxation of particle confined in more
general potentials. Here we consider the following form of the
potential (in units of $\beta^{-1}=1$)
\[ U(x)=
\begin{cases}
     \frac{1}{2}\left(x/L\right)^{\alpha}, \hspace{1cm} x \geq 0\\    
    \infty,         \hspace{2.2cm}x \leq 0
\end{cases}
\]
where $L$ should now be thought of a parameter of dimension length. As
previously there is Maxwell heat bath attached at the origin
$x=0$. Whenever a particle hits this reservoir it gets absorbed and
immediately comes out with a new velocity $u$ chosen from the Rayleigh 
distribution $f(u)$. The propagator is still given by the renewal equations
\eref{aconvolution} are \eref{mastereq-la}, where the functions on the
right hand side have to be computed for this case. It is clear
from \eref{mastereq-la} that the relaxation to the equilibrium is
determined mainly by the statistics of the returning time $\tau$ for
the particle to the reservoir, as we have previously noticed in the
box case. Therefore we focus on evaluating the return time
probability $I(\tau)$ defined before Eq. \eqref{aconvolution}.  The
expression of the time $\tau$ required by the particle to return back
to the reservoir in terms of it's initial velocity $u$ can be obtained
using the equation of motion, which gives
\begin{equation} \tau =2\int^{x_{\rm
max}}_{0}\frac{dx}{\sqrt{u^2-2U(x)}},~~~\text{where}~~~x_{max}=L~u^{2/\alpha}.\label{tau1} 
\end{equation}
This integral can be performed explicitly and gives
\begin{eqnarray}
 \tau= \mathcal{C}
 u^{\frac{2}{\alpha}-1}~~~~~\text{with}~~~~~~\mathcal{C}\equiv\mathcal{C}(\alpha, L)=  \frac{\Gamma(1+\frac{1}{\alpha})}{\Gamma(\frac{1}{2}+\frac{1}{\alpha})}\,
2\sqrt{\pi}\, L, \label{tau}
\end{eqnarray}
where $\Gamma(x)$ is the Gamma function.  Given the distribution
$f(u)=\theta(u)\, u e^{-u^2/2}$, the probability distribution
$I(\tau)$ of $\tau$ is obtained from $I(\tau)
=\int^{\infty}_0 \delta(\tau-\mathcal{C}
u^{\frac{2}{\alpha}-1})f\left(u\right)du$ which may be explicitly
written as
\begin{eqnarray}
 I(\tau)=\frac{\alpha}{|2-\alpha|}\left(\frac{\mathcal{C}}{\tau}\right)^{\frac{3\alpha-2}{\alpha-2}}\exp\left[-\frac{1}{2}\left(\frac{\mathcal{C}}{\tau}\right)^{\frac{2\alpha}{\alpha-2}}\right]
\end{eqnarray}
and its Laplace transform is given by:
\begin{eqnarray}
\tilde{I}(s) =\int^{\infty}_{0} e^{-s\mathcal{C} u^{(\frac{2}{\alpha}-1)}} f(u) du = \int^{\infty}_{0} u e^{-\frac{u^2}{2}-s\mathcal{C} u^{(\frac{2}{\alpha}-1)}} du.\label{a88}
\end{eqnarray}
It follows from \eref{mastereq-la} that the relaxation of the particle 
depends on the analytical properties of $\tilde{I}(s)$ in the complex 
$s$ plane.  To get some idea about this we look at the moments:
$\langle \tau^n\rangle=\frac{1}{n!}\frac{\partial^n \tilde{I}}{\partial
s^n}|_{s=0}$. For given $\alpha$, $\langle \tau^n\rangle$ diverges if
$\frac{n(\alpha-2)}{\alpha}\geq 2$ is satisfied. This implies that for
all $\alpha\leq 2$, $\langle \tau^n\rangle$ is finite for all $n$
while for $\alpha> 2$, $\langle \tau^n\rangle$ diverges for all
$n \geq \frac{2\alpha}{\alpha-2}$.  Interestingly, notice that the
bound reduces to $2$ as $\alpha \rightarrow \infty$ which is exactly
the box case studied in the previous section. $\alpha=2$ is very
special case where the $\tilde{I}(s)$ is independent of $s$.  This
suggests that the analytic properties of $\tilde{I}(s)$ is different
for different values of $\alpha$.

In order to obtain the type of relaxation,  we need to
look at the non-analyticities of $\tilde{J}(s)^{-1}$ in the complex
$s$ plane [see \eref{mastereq-la}], where
\begin{eqnarray}
\tilde{J}(s)\equiv1-\tilde{I}(s) =\int^{\infty}_{0} \left[1-e^{-s\mathcal{C} u^{\left(\frac{2}{\alpha}-1\right)}}\right] f(u) du. \label{a8}
\end{eqnarray}
Let us study  $\tilde{J}(s)$ for different values of $\alpha$. 

\begin{figure}[t]
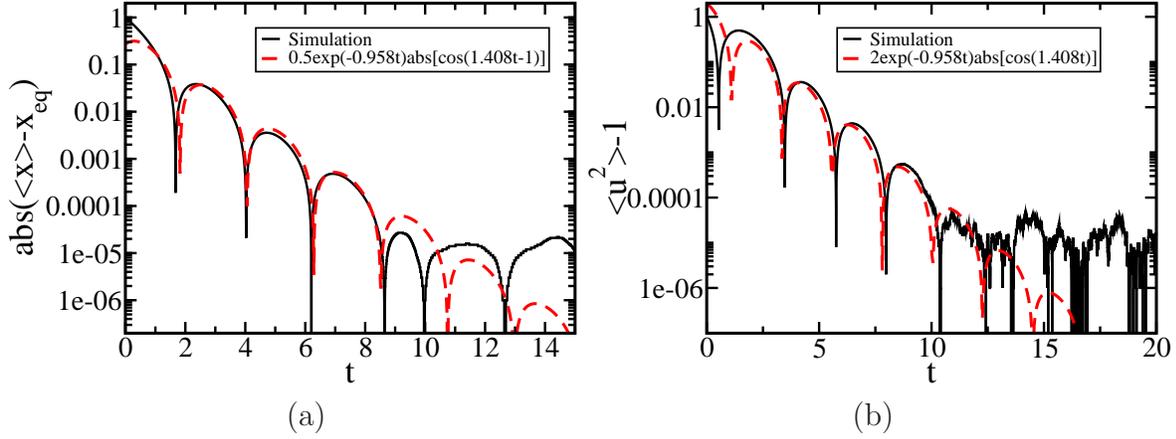

\centering
\includegraphics[height=5.1cm,keepaspectratio=true]{fig4a.eps}
\includegraphics[height=5.1cm,keepaspectratio=true]{fig4b.eps}\\
(a)\hspace{7cm}(b)
\caption{ Particle in a linear potential ($\alpha=1$): mean position $\langle x \rangle$ and mean squared velocity $\langle u^2 \rangle$ of the particle as a function of time obtained from simulations  is compared with analytical
prediction following from Eq.~(\ref{exprelax}).  The decay constant $\lambda_{re}$ and the oscillation frequency $\lambda_{im}$ are known exactly [see discussion after Eq.~(\ref{alpha=1})], while the  phase $\phi$ and amplitude  are obtained as fitting parameters.  Here $x_{eq}=1$ in the simulations in agreement with the expected exact result}
\label{alpha-1}
\end{figure}

\noindent
{\bf Case I:} $\alpha>2$.  The divergence of the moments, in this
case, suggests that the function $\tilde{J}(s)^{-1}$ has some other
divergences/non-analyticities apart from the $1/s$ divergence. So it
is a good idea to look at the behavior of $\tilde{J}(s)$ in the
$s \to 0$ limit, where the dominant terms are given by 

\begin{eqnarray}
\fl \qquad
\tilde{J}(s) &=&\int^{\infty}_{0} \left[1-e^{-s\mathcal{C} u^{-(\alpha-2)/\alpha}}\right] f(u) du =\sum^{\Delta}_{n=1}a_ns^n + b~ s^{2\alpha/(\alpha-2)}  ~ g(s,\alpha)+ \dotsb,~~~\label{J}
\end{eqnarray}
where $a_n=\frac{\mathcal{C}^n}{n!}\int u^{-n(\alpha-2)/\alpha}f(u)du$, $b$ is some constant, and 
\begin{equation}
\begin{cases}
g(s,\alpha)\simeq&\log(s)~~~~\text{and}~~~~ \Delta=(\alpha+2)/(\alpha-2),~~~~\text{if}~~2\alpha/(\alpha-2)~\text{is~an~integer}\\
g(s,\alpha)\simeq&1~~~~~~~~~~\text{and}~~~~ \Delta=\lfloor 2\alpha/(\alpha-2) \rfloor,~~~~~\text{otherwise}.
\end{cases}
\end{equation}
Here $\lfloor x \rfloor$ is the usual floor function indicating the integer part of $x$.
This behavior of $\tilde{J}(s)$ indicates that the relaxation to equilibrium happens as
\begin{eqnarray}
\fl
 &&\left [G(x,u,t|x_0,u_0) - P_{MB}(x,u) \right ]_{t \to \infty} \sim \frac{h(t)}{t^{(\alpha+2)/(\alpha-2)}},~~\text{where} \nonumber \\ 
 && \nonumber \\
 &&h(t)=
 \begin{cases}
 &\log(t),~~~~~~~~\text{if}~~2\alpha/(\alpha-2)~~\text{is~an~integer} \\
 &1,~~~~~~~~~~~~~~\text{otherwise}.
 \end{cases}
\end{eqnarray}
Note that, in the limit $\alpha \to \infty$, the approach to equilibrium is of the form $\left [G(x,u,t|x_0,u_0) - P_{MB}(x,u) \right ]_{t \to \infty} \sim \frac{\log(t)}{t}$ as shown for the particle in a box in the previous section.

\begin{figure}[t]
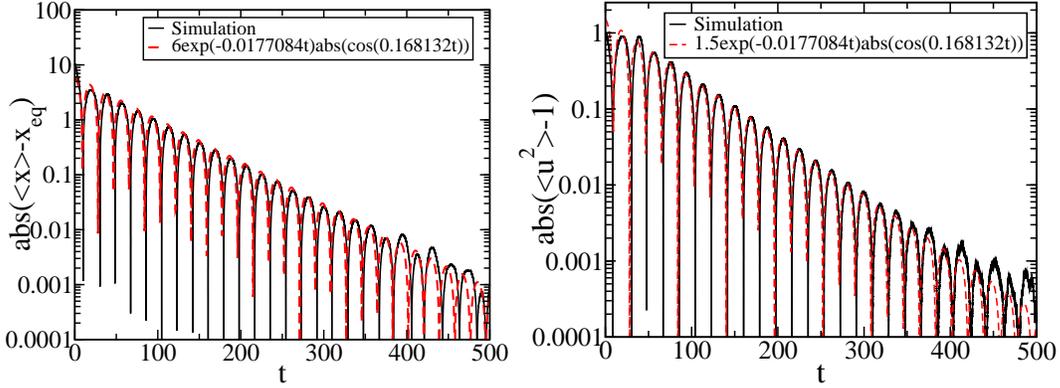

\centering
\includegraphics[height=5.1cm,keepaspectratio=true]{fig5a.eps}
\includegraphics[height=5.1cm,keepaspectratio=true]{fig5b.eps}
\caption{ Particle in a nonlinear potential with $\alpha=3/2$: mean position $\langle x \rangle$ and mean squared velocity $\langle u^2 \rangle$ of the particle as a function of time obtained from simulations  is compared with analytical prediction following from Eq.~(\ref{exprelax}).  The decay constant $\lambda_{re}$ and the oscillation frequency $\lambda_{im}$ are known exactly [see discussion after Eq.~(\ref{alpha=1})], while the  phase $\phi$ and amplitude  are obtained as fitting parameters. 
Here $x_{eq}=10.475$ in the simulations in agreement with the expected exact result} \label{alpha-1.5}
\end{figure}

\noindent
{\bf Case II:} $\alpha<2$. In this case, the function $\tilde{J}(s)^{-1}$ 
does not have any other divergences/non-analyticities apart from the $1/s$ 
divergence for real $s$. As a result the relaxation towards equilibrium, 
in this case is determined from the simple poles of $\tilde{J}(s)^{-1}$ 
[or zeros of $\tilde{J}(s)$] in the complex plane apart from the $s=0$ 
pole. The dominant contribution will come from the zero $\lambda= 
\lambda_{re}+ i~\lambda_{im}$ with largest $\lambda_{re} < 0$.  This 
suggests that, one has the usual exponential relaxation of the form
\begin{eqnarray}
\left [G(x,u,t|x_0,u_0) - P_{MB}(x,u) \right ]_{t \to \infty} \sim  \exp(-\lambda_{re}t) \cos(\lambda_{im}t +\phi) \label{exprelax}
\end{eqnarray}
where $\phi$ is phase factor which would arise depending on the initial 
condition (in our case, $\phi$ is obtained by fitting to simulation data). 
Interestingly, for the $\alpha=1$ case, explicit calculations can be done 
to find out the relaxation to equilibrium. In this case, the force acting on 
the particle $F=-\partial U/\partial x=-1/(2L)$ is a constant.
This implies $\mathcal{C}=4L$ 
and hence $\tau= \mathcal{C} u^{\frac{2}{\alpha}-1}=4 L u$. Therefore,
\begin{eqnarray}
\tilde{J}(s) &=&\int^{\infty}_{0} \left[1-e^{-4Ls u}\right] f(u) du =\sqrt{2\pi L}~s e^{\frac{(4Ls)^2}{2}} {\rm erfc}\left(2\sqrt{2} L s\right) \label{alpha=1}~.
\end{eqnarray}
Apart from the zero at $s=0$, the other zeros correspond to those of ${\rm
erfc}\left(2\sqrt{2} L s\right)$ \cite{Fettis,Zeros}. The zero at
$s=0$ corresponds to the equilibrium state while the  zero 
with largest negative real part determines the relaxation. The zero with largest real part has been computed numerically to high accuracy and   given by $2\sqrt{2}L s^*\approx -1.3548+ i 1.9915$ \cite{Fettis,Zeros}. For  $L=0.5$ (corresponds to $F=-1$), $\lambda_{re} \approx 0.958$ and $\lambda_{im}=1.408$. For other values of $\alpha<2$, one can follow the same procedure and study the relaxation.  For $\alpha=3/2$, $L=10$, we find,  from a numerical evaluation of the zeros of $\tilde{J}(s)$, $\lambda_{re}\approx -0.0177084 $ and $\lambda_{im} \approx  0.168132$. In Fig. \ref{alpha-1} and  Fig. \ref{alpha-1.5}, we compare these analytic predictions with simulation results for the relaxation of $\langle x \rangle$ and $\langle u^2 \ra $ to their equilibrium values, for the cases $\alpha=1$ and $\alpha=3/2$ respectively. In both cases we find good agreement between the simulated and analytical results.

\noindent
{\bf Case III:} $\alpha=2$: In this case the particle does not
equilibrate at all, even in the long times, because the time period of
a harmonic oscillator does not depend on the initial velocity (it
depends only on the stiffness of the potential and the particle
mass). The required $\tau=\pi L$ [see Eq. \eqref{tau}] by the particle
to return back to the reservoir is same for all $u$ with which it
started.  Therefore, unlike the previous case, the revisiting time of
the particle is not a random quantity.  As a result after every time
interval $\tau=\pi L$, phase-space density again becomes a delta
function, consequently the system does not show enough mixing even in
the sufficiently long time and therefore fails to equilibrate.

\section{Discussion}
\label{Conclusion}
The dynamics of a particle in a confining potential in the presence of
a thermal environment is often described by the Langevin equation
which consists of a dissipative part and a noise which constantly act
on the particle. For a large class of potentials it can be shown that
the relaxation to equilibrium takes place exponentially fast in time
and this is related to the fact that the Fokker-Planck operator
corresponding to the Langevin dynamics has a discrete spectrum.
However, in many situations, the heat bath could be localized in space
and it effects the particle's motion only when the particle is located
in a particular spatial region. We address the question of relaxation
to equilibrium for such localized baths.

Here we considered a single particle moving on the positive half line
in a confining potential [$V(x)\sim x^\alpha$], with a hot hard wall
placed at the origin. Particles that hit the wall emerge with a new
velocity, chosen from the Rayleigh distribution (this ensures thermal
equilibration at long times). Some special choices of the potential
correspond to particle in a box with reflecting walls at the right
end, a particle in a gravitational field and a particle in a harmonic
potential. We find that the relaxation to thermal equilibrium can
occur in a variety of ways. Depending on the form of the potential,
one can have exponential relaxation (for $\alpha <2$) or power-law
relaxation ($\alpha >2$), sometimes with a logarithmic correction.
Interestingly, for $\alpha=2$, the system fails to equilibrate even in
the long time limit.  The origin of the slow power-law relaxation can
be understood as arising from the heavy-tailed distributions for the
time taken by the particle emerging from the bath (at the origin) to
return to the origin.  The statistics of the returning times plays a
crucial role, because, in the approach towards equilibrium, the
particle has to repeatedly interact with the bath.  In our study of
the relaxational dynamics we used a renewal approach. For the special
case of the particle in a finite box, we also obtained an exact
solution of the master equation describing the evolution of the
probability distribution.

The present study may be taken forward in several other
directions. Non-thermal reservoirs would be of interest in the context
of approach to non-equilibrium steady-states of systems such as a ball
bouncing on a vibrated plate \cite{Chastaing,Luck}. For example, a 
non-equilibrium reservoir, where the velocity emerging from
the reservoir has a power-law distribution, would give rise to a
power-law tail for return time even for $\alpha <2$, and hence, would
result in slow relaxation to the steady state. Another interesting
case is that of heat transport in low-dimensional systems where a
popular model consists of a gas of hard particles in a one-dimensional
box with localized thermal walls at different temperatures at the two
ends \cite{AD}. Again the system reaches a non-equilibrium steady state
and an important question is the nature of approach to the steady
state.

\section{Acknowledgements} 
We acknowledge support from the Indo-French Centre for the Promotion of Advanced Research (IFCPAR) under project 5604-2. AD would like to thank support from the Indo-Israel joint research project No. 6-8/2014(IC) and from the grant EDNHS ANR-14-CE25-0011 of the French National Research Agency (ANR).


\end{document}